\documentclass[12pt]{iopart}

\usepackage{iopams}  
\usepackage{graphics}
\usepackage{amssymb}
\usepackage[jphysicsB,round]{harvard}

\begin{document}

\title{FIRST: Fast Iterative Reconstruction Software for (PET) Tomography}
\author{J L Herraiz$^1$, S
 Espa\~na$^1$,
J J Vaquero$^2$,
M Desco$^2$ and
J M  Ud\'{\i}as$^1$}
\address{$^1$ Dpto. F\'{\i}sica At\'omica, Molecular y Nuclear, Universidad Complutense de Madrid, Spain}
\ead{jose@nuc2.fis.ucm.es}
\address{$^2$ Unidad de Medicina y Cirug\'{\i}a Experimental, Hospital GU Gregorio Mara\~n\'on, Madrid, Spain}

\begin{abstract}
Small animal PET scanners require high spatial resolution and good sensitivity. 
To reconstruct high resolution images in 3D-PET, iterative methods, such as OSEM, 
are superior to analytical reconstruction algorithms, although their high computational 
cost is still a serious drawback. The higher performance of modern computers could 
make iterative image reconstruction fast enough to be viable, provided we are able 
to deal with the large number of probability coefficients for the system response 
matrix in high-resolution PET scanners, which is a difficult task that prevents the 
algorithms from reaching peak computing performance. Considering all possible axial 
and in-plane symmetries, as well as certain quasi-symmetries, we have been able to 
reduce the memory requirements to store the system response
matrix (SRM) well below  1 GB, which allows us to 
keep the whole response matrix of the system inside RAM of ordinary industry-standard
 computers, so that the reconstruction algorithm can achieve near peak performance.
The elements of the SRM are stored as cubic spline profiles and matched to voxel size
 during reconstruction.
In this way, the advantages of `on the fly' calculation and of fully stored SRM are combined.
 The on the fly 
part of the calculation 
(matching the profile functions to voxel size) 
of the
SRM 
accounts for 10\% to 30\% of the reconstruction time, depending on the number of voxels chosen. 
We tested our approach with real data from a commercial small animal PET scanner. The results 
(image quality and reconstruction time) show that the proposed technique is a feasible solution.

\end{abstract}



\maketitle

\section{Introduction}

There is a strong demand for fast and accurate reconstruction procedures for high-resolution, high-sensitivity
PET scanners. These systems, typically used in small animal PET studies, are designed with the goal of
optimizing spatial resolution while maintaining good detection sensitivity. They consist of several pairs of opposite
scintillation detectors, each of which is coupled to an array of small crystals arranged in a static ring or in
a rotating device. The detector ring diameter and the size of the field of view (FOV) are usually less than 20 cm.
Since spatial resolutions in the range of 1 mm are required, these detectors employ many small 
crystals, thus leading to a very large number of lines of response (LORs),  defined by every possible pair
of crystals. Moreover, 3D acquisitions (and reconstructions) are mandatory due to sensitivity requirements. 

Statistical 3D reconstruction methods such as Expectation Maximization (EM) \cite{Shepp,Browne} have shown superior
image quality than conventional analytic reconstruction techniques. Moreover, EM has some desirable properties
such as conservation of the number of counts, non-negativity, good linearity and dynamic range. 
One of the key advantages of statistical reconstructions is the ability to incorporate accurate models of the
PET acquisition process through the use of the system response matrix (SRM). However, SRM for 3D systems are
of the order of several billions of elements, which imposes serious demands for statistical iterative
 methods in terms of the time required to complete the reconstruction procedure and the computer memory
 needed for the storage of the SRM. 
We have designed, developed, implemented and tested a new EM-based reconstruction methodology that provides fast
reconstructions while remaining very flexible. With our approach, the usual difficulties of iterative reconstruction
methods regarding the large size of the SRM or the use of unrealistic estimates of it, are overcome by means 
of a compressed and realistic SRM. The efficiency of the proposed method relies on the design and method of 
storing this SRM. The imaging process of obtaining the $y(i)$ counts on each of the $i$ pair of detectors, from an
object discretized in $x(j)$ voxels  can be described by the operation 
$y(i)=\sum_j A(i,j) x(j) $
where $A(i,j)$ is the SRM, the vector $x(j)$ corresponds to the voxelized
 image and $y(i)$ to the measured data. 
Each element ${A(i,j)}$ is defined as the probability of detecting an annihilation event coming from image voxel 
$j$ by a detector pair $i$. This probability depends on factors such as the solid angle subtended by the
voxel to the detector element, the attenuation and scatter in the source volume and the detector response
characteristics.

The forward projection  operation just introduced above,
estimates the projection data from a given activity distribution of the source. Backward projection is 
the transposed operation of forward projection; it  estimates a source volume distribution of activity from the
projection data. The operation corresponds to  
$b(j)=\sum_i A(i,j)y(i)$ 
where $b(j)$ denotes an element
of the backward projection image.
Both the forward and backward projection operations require knowledge of the SRM \cite{Frese,Rafecas}.
Iterative reconstruction algorithms repeatedly use the forward and backward projection operations, which are
the most time-consuming parts of iterative reconstruction programs. Some implementations trade accuracy for
speed by making approximations that neglect some physical processes, such as positron range, scatter and 
fractional energy collection at the scintillators, or visible light loses 
in the detectors \cite{Vaquero1,Vaquero2,Yamaga}. This approach simplifies these operations to increase speed, 
but this trade off often leads to non-optimal images. 

The evaluation and storage of SRM elements is a very active subject of research. Ideally, the SRM could be
calculated, using MC methods \cite{Rafecas} or from empirical data \cite{Frese}, and stored once and for all 
before the beginning of the reconstruction process. In practice, memory requirements for doing this
have become prohibitive so far. 
A number of methods have been proposed to compute and handle huge sparse matrices like the SRM. Some implementations
compute the elements ${A(i,j)}$ on the fly, only if and when they are required \cite{Kudrolli}, thus avoiding 
the need to store the whole SRM. In other approaches, the SRM has been factorized as a product of independent 
contributions: geometry, attenuation, and detector sensitivity \cite{Qi}. The simplifications required by on-the-fly
or factorized calculations often overlook important effects \cite{Lee}. 

Due to the fact that the SRM is very large but sparse, it may be kept on disk by using sophisticated storage schemes 
and taking advantage of system symmetries \cite{Calvin} to reduce the size of the SRM to a few tenths of
Gigabytes. Due to the fact that accessing the SRM from disk for every forward and backward projection operation is very slow, this considerably slows down the reconstruction.
Our approach involves compressing the SRM to the extent that enables its allocation in the RAM memory of industry
standard computers, avoiding disk access during reconstruction. In this way, it is possible to achieve a
sustained performance of around 50\% of the theoretical peak computing capability of the processors.

\section{System Response Matrix  [SRM]}

The SRM is composed of all the $V \times  L$ probability elements  $A(i,j)$, $i=1,\cdots,L$, $j=1,\cdots,V$ 
representing the probability of detecting an event coming from voxel $V(j)$ at detector LOR (Line of Response)
 $L(i)$. Forward and backward projection require the knowledge of all of these elements. This matrix depends
 on factors such as the physics of beta decay, attenuation and scatter in the source volume, solid angle subtended 
from voxel to detector element, and intrinsic detector response characteristics. For a reconstruction method 
to be accurate, all these effects should be considered.
The equipment used in this study is an eXplore Vista-DR (GE) small animal PET scanner \cite{Vaquero1}. It is
 a ring-type scanner with a diameter of 11.8 cm, a transverse Field of View (FOV) of 6.8 cm and an axial FOV of 
4.6 cm.
 Vista technology is based on scintillator detector modules with depth-of-interaction (DOI) capabilities
 \cite{Vaquero2} arranged in single (SR) or double rings (DR). The detector modules are composed of a
 13 $\times$ 13-crystal
 array with 1.55 mm pitch size. The number of LORs in this scanner is over 2.8$\times 10^7$ (see Table \ref{tb:tb1}). DOI 
determination enables spatial resolution and sensitivity to be improved simultaneously \cite{Yamaga}.

\begin{table}
\caption{\label{tb:tb1}VISTA PARAMETERS}
\begin{indented}
\item[]\begin{tabular}{@{}*{7}{l}}
\br
Ring diameter&11.8 cm\\
Aperture&8 cm\\
Axial FOV&4.8 cm\\
Number of DOI detector modules:&36 position-sensitive PMTs\\
Number of dual-scintillator DOI elements&6,084\\
Crystal array pitch&1.55 mm\\
Total number of crystals&12,168\\
Total number of 3D coincidence lines& $28.8 \times 10^6$ \\
\br
\end{tabular}
\end{indented}
\end{table}


%
        From the data of Table \ref{tb:tb1} and at nominal image resolution of 175 $\times$ 175 $\times$ 62 voxels
(near 1.9 millions of voxels) the number of elements in the SRM (number of LORs $\times$ number of voxels) 
is of the order of $5 \times 10^{13}$.
Storing all the elements of the SRM would require more than 10 TB \cite{Rafecas}. This exceeds the resources 
of any ordinary workstation, making it necessary to disregard all redundant elements and to perform 
approximations in order to be able to store the SRM in the limited amount of RAM of ordinary workstations. Three 
techniques have been used to achieve
 this goal: Null or almost-null element removal (matrix sparseness); intensive use of system
 symmetries; and compression of the resulting SRM employing quasi-symmetries.

\subsection{Matrix sparseness}
Every detector pair can receive coincidence counts only from a relatively small portion of the FOV. 
Thus,  most matrix elements of the SRM are null and only the non-zero elements should be stored, 
reducing considerably the storage requirements. To estimate how many non-zero elements 
of the SRM have to be taken into account, we proceed as follow: The voxels
 connected to a given LOR (that is, the voxels from which positron decay can produce 
with non negligible probability 
 a valid coincidence count 
in the detectors that define the LOR),
  constitute the so-called 
``channel of response'' or CHOR \cite{Michel} for that LOR.
Extensive simulations determine the maximum size of the CHOR needed and only the SRM
elements that are part of some CHOR are stored.
We consider a voxel not connected to a LOR ({\em i.e.}, not being part of the CHOR) if  the probability that
a positron emitted from that voxel yields a count in the corresponding detector pair  is smaller than
 one thousandth of the maximum value of all the voxels for such  given CHOR. For 
the scanner considered here
using the nominal number of voxels  of 175 $\times$ 175 $\times$ 62 in XYZ to cover the FOV, which
yields a voxel size of $0.38 \times  0.38 \times 0.76$ mm$^3$ 
(see Table \ref{tb:tb1}) and an
  average
 number of voxels in a CHOR of about 6000 for a typical CHOR size of 150 (longitudinal) $\times$ 
10 (transverse width  of the CHOR in the transaxial XY plane of the scanner) $\times$
  4 (transverse width  in the axial or Z direction of the scanner) voxels  \cite{Calvin}. 
With this choices, the number of nonzero elements of the SRM is then 28.8 
$\times$ $10^6$ LORs $\times $ 6000 connected voxels on average, or  $10^{11}$ elements. 
That is, only around 0.2\% of the elements 
of the SRM are nonzero. Yet, storing these nonzero elements as floats (4 bytes  per SRM element) will
require about 600 GB of disk space,  still too high for the current
RAM amount of industry-standard computers.

\subsection{System Symmetries}

\begin{figure}
\begin{center}
\rotatebox{270}{\scalebox{0.6}{\includegraphics{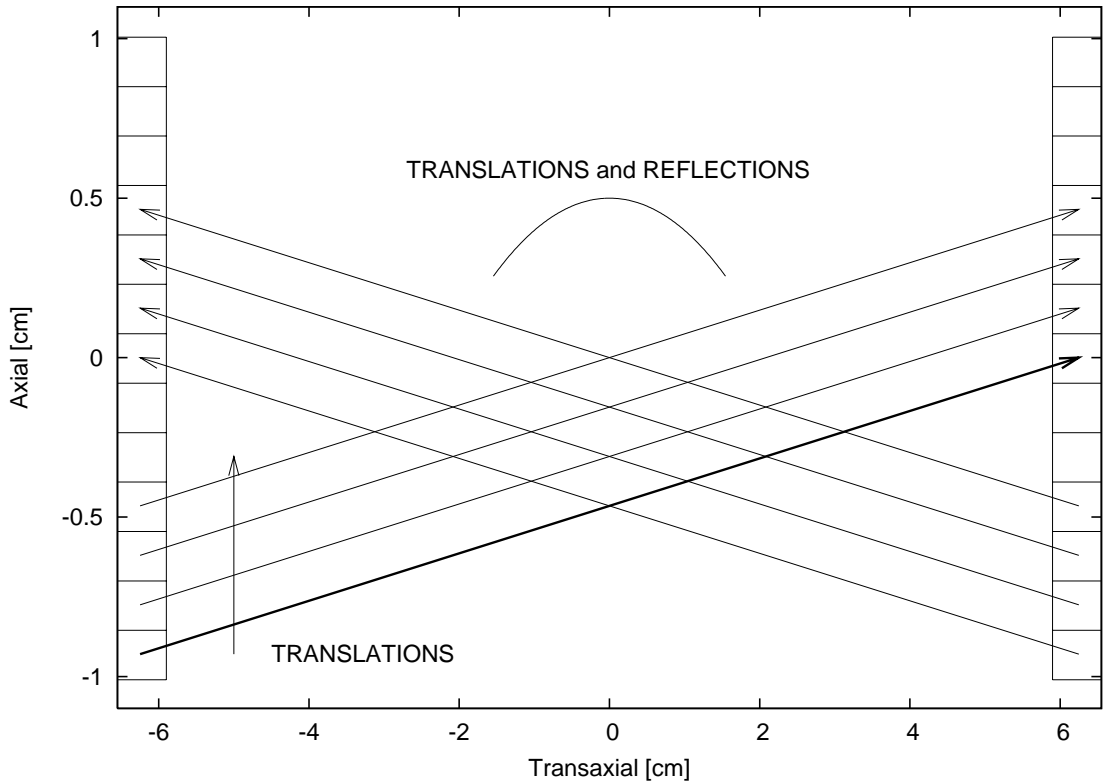}}}
\caption{Schematic Drawing of a VISTA small animal PET scanner detector pair, showing
the [exact] translation and reflection symmetries employed in this work. All the
elements of the SRM belonging to parallel LORs are, by symmetry, equivalent.} \label{fig:fig1}
\end{center}
\end{figure}

An additional reduction factor of approximately 40 in the number of non-null SRM elements that need to be stored
 can be achieved by assuming that [exact] axial (translation and reflection) and in-plane symmetries exist
 \cite{Calvin}. Voxels were chosen in an orthogonal grid oriented along X,Y and Z axis, Z being the
axis of the scanner.  If an integer number of voxels is employed for the width of the CHOR in the Z axis, then 
there is a Z-translation symmetry, due to the fact that voxels in the same relative position of the CHOR and 
belonging to parallel CHORS should have equal values (see Fig. \ref{fig:fig1}).

 We must however note that, although our SRM  exhibits indeed this translational symmetry,
 in real scanners, due to edge and block effects, it is only an approximate symmetry.

There is also another axial symmetry, Z-Reflection symmetry. Using both parallel and reflection Z-symmetries,
 the number of elements to be stored is reduced considerably. Each pair of blocks has $(2\times 13\times 13) 
\times (2\times 13\times 13)$ LORs,  but using symmetries, only $ 2\times 13 \times (2\times 13\times 13)$ 
need to be stored. A factor 13 of reduction in the space needed is achieved.

Another symmetry, reflection symmetry among blocks in the XY plane, also holds.
 Using this, the number of pairs of detectors that have CHORs with different values is reduced by  a factor of  3.
Using them all, as we have already mentioned, these symmetries allow to reduce by a factor or near 40 the number of 
different elements of the SRM that must be stored.
Storage needs can thus be reduced to a few  (slightly less than 10, for the scanner we consider here) GB, small 
enough to fit in hard disks, yet too much to be maintained in RAM. 

\subsection{Compressed SRM}
The last step of the method we propose here uses additional non-exact symmetries, or 
{\em quasi-symmetries}, in order to obtain additional reduction of the SRM.
If we allow for relatively small differences between quasi-symmetric elements of the
 SRM ({\em versus} no difference {\em a priori} in the case of the exact symmetries), we can group certain LORs 
 into sets of the same quasi-symmetry class. The differences between the elements
 of the SRM for LORs belonging to a given class should be much smaller than between LORs from different
 classes. Quasi-symmetry classes can be obtained, for instance, by grouping together LORs from
 crystals with different, but close, LOR-crystal orientations. The differences among the elements of the
 same quasi-symmetry class are about 5\% to 10\%, depending on the amount of compression 
(reduction
in size) desired.

\begin{figure}
\begin{center}
\rotatebox{270}{\scalebox{0.6}{\includegraphics{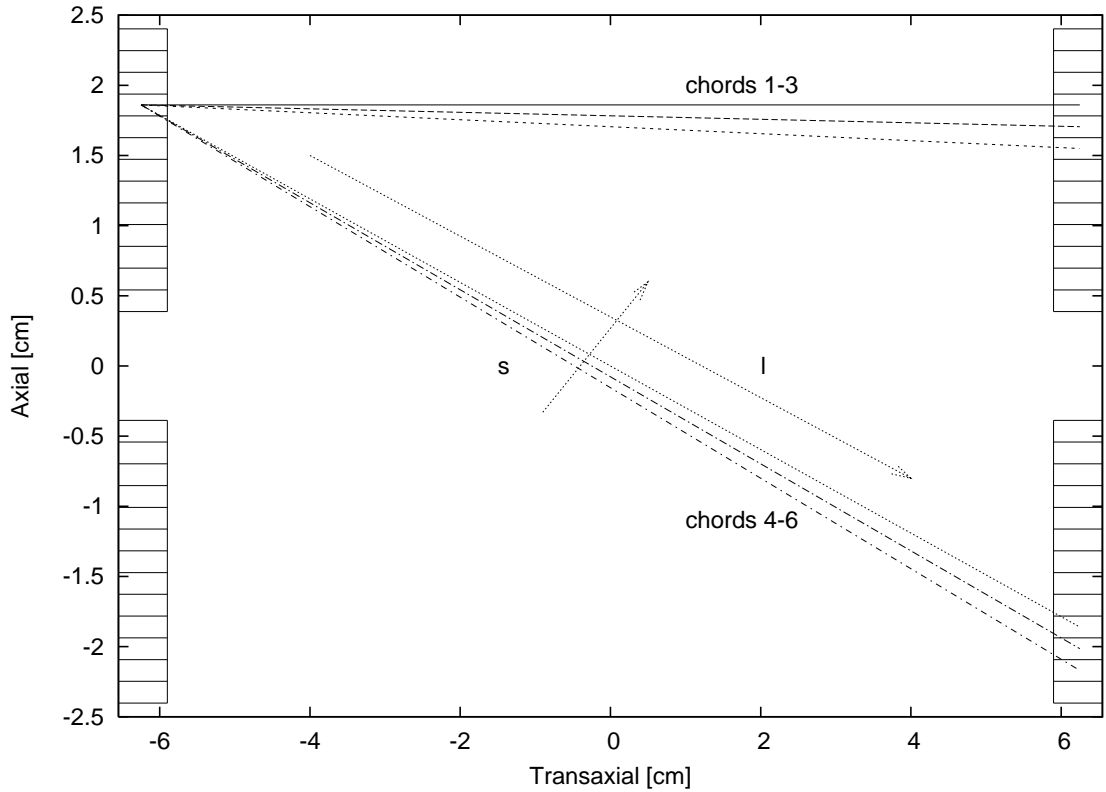}}}
\caption{Schematic representation of several lines of response (LOR) considered
for the discussion on quasi-symmetries. Three LORs (numbered 1 to 3 from top to bottom) with a 
small relative LOR-crystal angle and three (numbered 4 to 6, also from top to bottom) with 
large relative LOR-crystal angle are depicted. $l$ and $s$ are
the coordinates along the LOR direction and normal to it, respectively.
} \label{fig:chords3}
\end{center}
\end{figure}

\begin{figure}
\begin{center}
\rotatebox{270}{\scalebox{0.6}{\includegraphics{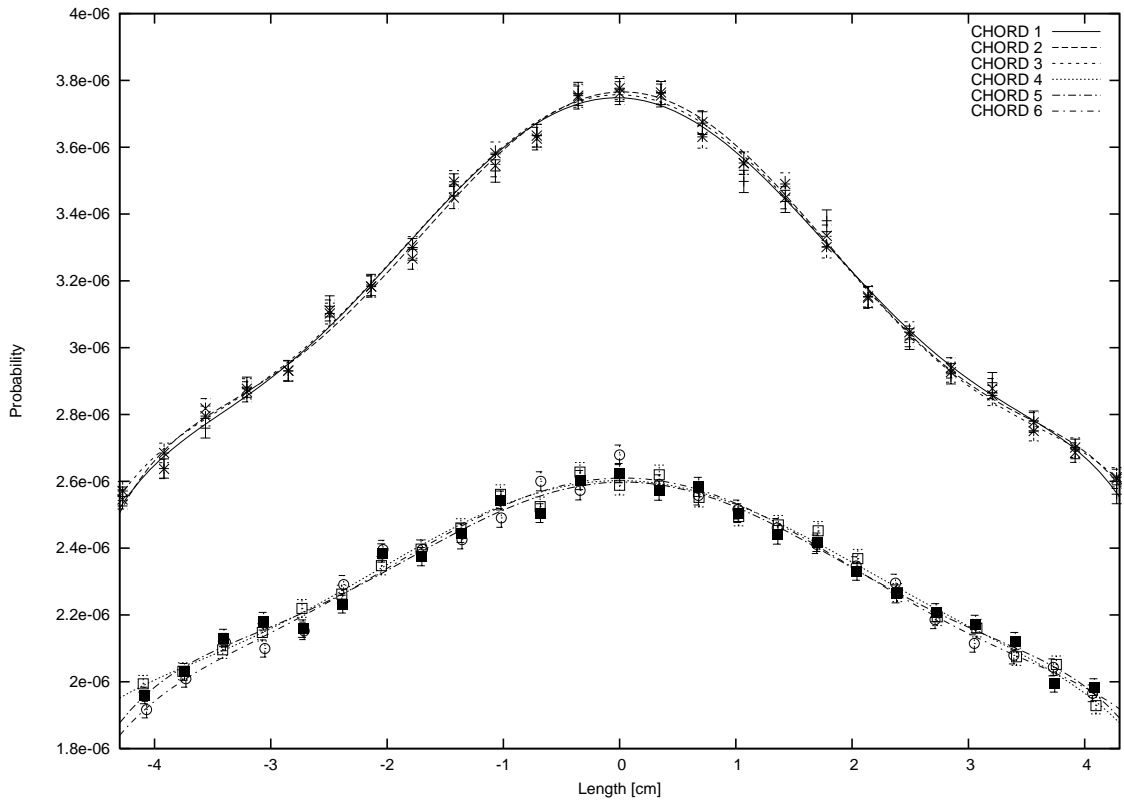}}}
\caption{Longitudinal profile of the probability elements for the  LORs shown in
figure \protect{\ref{fig:chords3}}. The probability of detection of a coincidence count in LORs 1 to 6 per
every positron decay in the axis of the CHOR as a function of the distance to the center of the CHOR is shown. The data points represent the results for the MC simulation described in the text, the error bars given by the statistical uncertainty. Symbols employed are as follows, 1:  plus ($+$) sign, 2:  times ($\times$) sign, 3:  star (*) sign, 4: empty square, 5: solid square,
6: empty circle. Profiles fitted to the points by means of cubic splines are also shown. 
The small crystal-LOR angle (1 to 3) profiles are very similar among them, but rather different from 
the large crystal-LOR angle ones (4 to 6). } 
 \label{fig:chords_long}
\end{center}
\end{figure}

\begin{figure}
\begin{center}
\rotatebox{270}{\scalebox{0.6}{\includegraphics{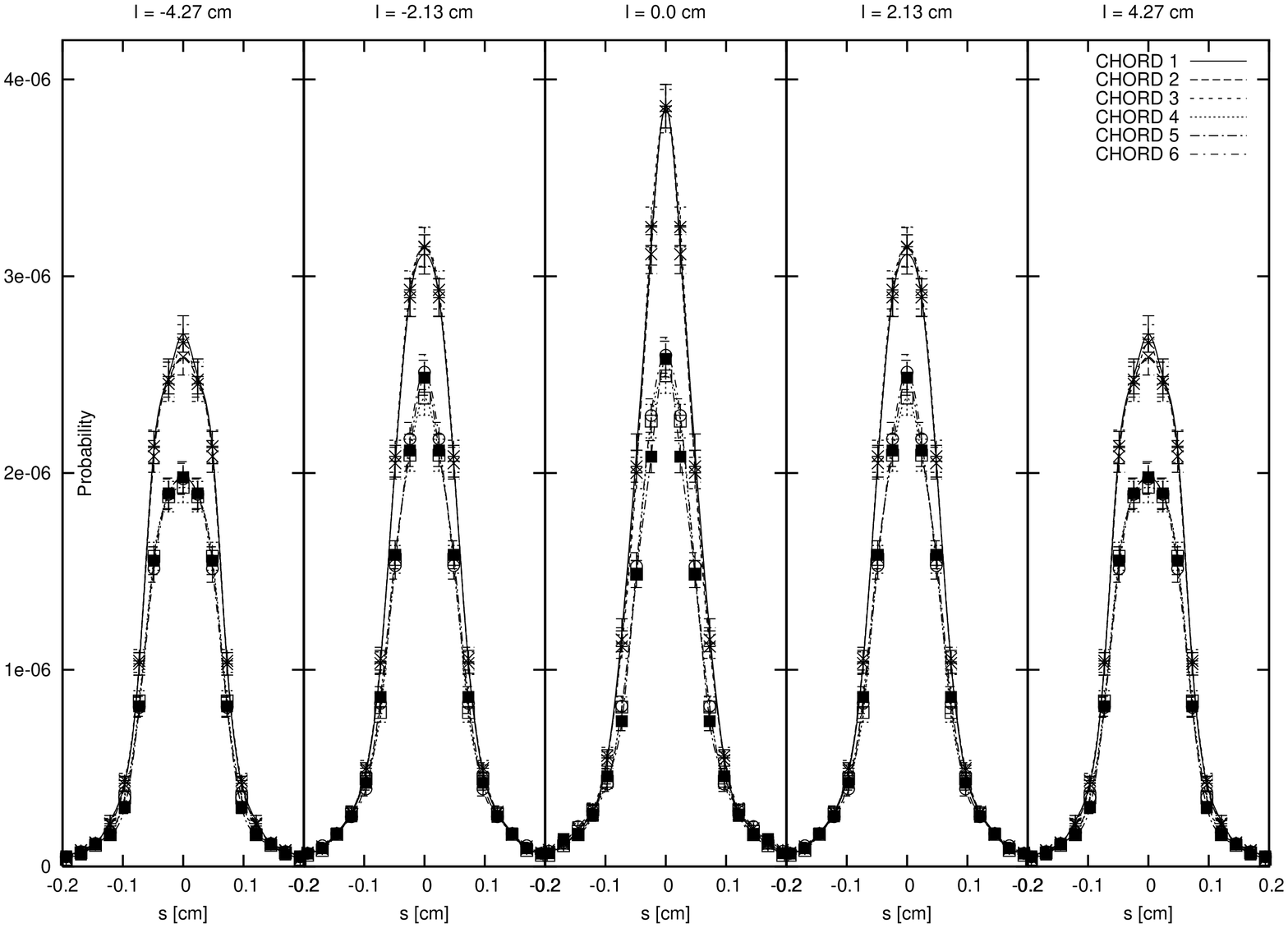}}}
\caption{Transverse profiles of LORs 1 to 6 of Fig. \protect\ref{fig:chords3}. 
Data points and curves as in previous Fig. \protect\ref{fig:chords_long}. Several profiles at different
distances l to the center of the LOR are shown. The transversal width of the  CHORs as shown in the figures is
around 4 mm, or three crystal widths. 
} \label{fig:chords_trans}
\end{center}
\end{figure}

In Figures \ref{fig:chords3}, \ref{fig:chords_long} and \ref{fig:chords_trans} we illustrate this procedure with an example taken from our simulations. MC
events were generated at different positions inside a CHOR. As shown in Figure \ref{fig:chords3}, LORs 1 to 3
are parallel or almost parallel to the crystals and thus the probability values along these three CHORs
should be very similar. Analogously, LORs 4 to 6 have a large LOR-crystal angle, similar for the three
of them. In Figure \ref{fig:chords_long}, longitudinal profiles along these LORs are shown. We can see indeed that the result of 
the MC simulations for the calculation of the
probabilities for LORs 1 to 3 (and 4 to 6), shown by the data points 
(that include statistical error bars), are very similar. We could even say that are the same within the error bars. 
In Figure \ref{fig:chords_trans}, now the profiles of the CHORs along the transverse direction to the LORs (s-coordinate)
are shown at several values of the l-coordinate. As in the case of the longitudinal profiles, we can
see that the results of the simulated data for the near equivalent LORs 1 to 3 or 4 to 6 are very similar. 
Also we realize that the variation of probability inside a CHOR is smooth, which allows us to 
fit the simulated points to profiles with an smooth interpolating cubic spline. 
The differences among the results of the interpolating curves for LORs 1 to 3 is marginal and the three
interpolating curves could be considered identical within the statistical error bars of the MC points. A similar
observation can be made for CHORs 4 to 6. Our quasi-symmetry assumption means that we will employ the
same profile functions for CHORs 1 to 3, that belong to the same quasi-symmetry class. CHORs 4 to 6
belong to another quasi-symmetry class and be represented by one set  of profile
functions, different from the ones of the other quasi-symmetry classes. 

  Depending on the geometry of the system, the use of quasi-exact symmetries allows us to obtain
 a number of quasi-equivalent LOR classes (that is, CHORs with non equivalent values) which may be 
 9 (in the example so far discussed) to 25 (for up to 5 different LOR-crystal angles in the
same quasi-symmetry class, allowing for larger differences among the profiles 
within the same quasi-symmetry class) times smaller than the number of classes obtained with
only exact symmetries. The elements of this notably reduced SRM are encoded as  
transverse and longitudinal profile functions obtained by cubic spline interpolation of MC sampled points. 
For each transversal or longitudinal profile, MC estimates of probability at $20 $ 
points along or across the CHORs  are employed to determine the cubic spline profile functions. 
At reconstruction time,  the probability element of the SRM corresponding to each voxel 
inside a CHOR is obtained by interpolation of the profile functions.  
If the voxel size chosen is large, we average several values interpolated from the profile functions 
at different points inside the voxel,  in order to compute the probability for each voxel.
The interpolation and averaging of probability inside each voxel from profile functions is
compared with the results of averaging of points obtained directly from the MC simulations. From these
comparisons, we conclude that for a number of voxels above or below a factor of three of the
nominal number of voxels of $ 175 \times 175 \times 62 $,  the interpolation procedure 
differs typically by less than 5\%  from the results of direct MC simulation in the example shown in Figures \ref{fig:chords3}, \ref{fig:chords_long} and \ref{fig:chords_trans}
(3 different LOR-crystal angles) and by less than 10\% for larger quasi-symmetry class (5 different
LOR-crystal angles).

 In short,  
the same quasi-equivalent profiles  can be used to build the non-zero SRM elements for a reasonable range 
of voxel sizes. The optimal voxel size that should be employed for each reconstruction may be different
depending for instance on the number of counts of a particular acquisition. 
  This profile function approach makes it possible to generate reconstructions 
with different voxel size without the need to recompute the SRM elements.
Eventually, this process leads to a compressed SRM that fits in 30 to 150 MB, depending
on the degree of quasi-symmetry assumed.

Toio end this section on compressing the SRM, we comment on the storing strategy that is also useful
 to save space.
All the cubic spline coefficients for the profiles of a quasi-symmetry class (or superCHOR),  
are rescaled in order to convert (and store) them as   integer values. 
Two bytes are employed to represent every coefficient, which allows 
to represent ratios of more than 60,000 to 1 inside the same CHOR.
The scale factors
(maximum and minimum values of the coefficients for all the profiles in each superCHOR) are also 
recorded as 
two additional floating numbers. 
During reconstruction, the integer values are converted into the adequate float 
ones on the fly.
 The FOV is divided in voxels arranged in an orthogonal
 grid. For a given CHOR, voxels are visited from bottom to top, 
left to right and back to front directions.  Every voxel in the superCHOR is visited in order and the SRM element
corresponding to that voxel-LOR combination is obtained by interpolation of the cubic 
spline profiles. Then, the superCHOR
values are stored as a list of  numbers formed by 
the probabilities of each voxel of the superCHOR visited in the assumed order. Once the superCHOR is
obtained (decompressed) on the fly, all the operations (forward or backward projections) that involve the CHORs
in the quasi-symmetry class are performed.
%

\subsection{MC simulation}
Given the fact that the compressed SRM fits in RAM, it does not need to be computed during 
reconstruction, nor read from disk once loaded in memory at the beginning of the reconstruction. 
Thus, the SRM can be computed using a very realistic model, and stored once and for all. 
Monte Carlo (MC) methods are, in principle, well suited to provide realistic estimates of 
SRM elements. In our case, we use our own MC model that includes scatter and incomplete
collection of energy in the scintillator crystals, positron range and 
non-colinearity effects. Positron range is dependent on the object. In order to incorporate 
its effect in the SRM,  in our simulations
the range is computed assuming  that water fills uniformly the whole FOV.  
Most importantly, we also include the scatter of gamma photons when 
they reach the scintillator crystals. The comparison of simulated phantoms with  actual 
acquisitions reveals that the simulation is very realistic indeed.

A large number of simulated events are accumulated until the statistical uncertainty is below 
5\% at the center of a typical LOR. Several weeks of computing time were required for the 
calculation of the SRM in a cluster composed of 12 industry-standard workstations. The total 
time employed for the full MC simulation is equivalent to 180 days of a single Pentium IV 3.0 GHz 
workstation.

\section{ITERATIVE  IMAGE RECONSTRUCTION ALGORITHMS} 

To test the accuracy of the compressed SRM we have obtained, we 
use one of the most widely applied algorithm for finding the maximum-likelihood (ML) estimation 
of activity $x$ given the projections $y$, that is expectation-maximization (EM). This was first 
applied to the emission tomography problem by Shepp and Vardi \cite{Shepp}. ML, though, is a 
general statistical method, formulated as a method of solving many different optimization problems.

Usually, iterative algorithms obtained from the ML statistical model assume that the data being 
reconstructed retain Poisson statistics. However, to preserve the Poisson statistical nature of data 
it is necessary not to perform any pre-corrections \cite{Qi}. Corrections for randoms, scatter and 
any other effects should be incorporated into the reconstruction procedure itself, rather than being 
applied as pre-corrections. At times, sophisticated rebinning strategies are
employed to build  sinograms  into radial and angular sets, what changes the 
statistical distribution of the data, which may no longer be Poisson-like \cite{Kadrmas}. Furthermore, 
much attention must be paid in order that sinogram rebinning does not cause a loss in the potential 
accuracy of the reconstruction. Our approach does not involve sinograms in any way, thus preserving 
all the information obtained during the acquisition. Uncorrected data, binned into raw 3D-LOR 
histograms, should  maintain Poisson statistics \cite{Kadrmas}.

A serious disadvantage of the EM procedure is its  slow convergence \cite{Lewitt}. This is due to 
the fact that the image is updated only after a full iteration is finished, which implies that 
all the LORs have been projected and back-projected at least once. In the Ordered Subset EM (OSEM) 
algorithm, proposed by Hudson and Larkin \cite{Hudson}, the image is updated more often, which 
has been shown to reduce the number of necessary iterations to achieve a convergence equivalent 
to that of EM.

According to the literature, EM methods have another important drawback: Noisy images are obtained 
from over-iterated reconstructions, and this is usually attributed to either the fact that there is 
no stopping rule in this kind of iterative reconstruction, or to the statistical (noisy) nature of 
the detection process and reconstruction method \cite{Bettinardi,Lagendijk}. In practice, however, 
an image of reasonable quality is obtained after a few iterations. 

Several techniques have been proposed to address the issue of the noisy nature of the data: Filtering 
the image either after completion of the reconstruction, during iterations or between them \cite{Slijpen}, 
removal of  noise from the data using wavelet-based methods \cite{Mair}, or smoothing the image with  
 Gaussian kernels (Sieves method) \cite{Snyder,Liow}.

Maximum A Priori (MAP) algorithms are also widely used \cite{Green}. MAP adds a priori information 
during the reconstruction process, the typical assumption being that due to the inherent finite 
resolution of the system, the reconstructed image should not have abrupt edges. Thus MAP methods 
apply a penalty function to those voxels which differ too much from their neighbors. Whether the 
maximum effective resolution achievable is limited, or even reduced, by the use of these methods 
is still an open issue.
On the other hand, a proper choice of the reconstruction parameters such as the number of iterations,
the use of an adequate system response and a smart choice of subsetting, allows high quality images
to be obtained by the EM procedure.

We implemented an OSEM algorithm that includes the possibility of MAP by means of 
a generalized one-step late MAP-OSEM algorithm, similar to the one described in 
\cite{Lewitt,Kadrmas}:

\begin{equation}
x^{n,s+1}(j)=x^{n,s}(j)·\left[\frac{\displaystyle\sum_{i\in Subseti\,\,\, S}{A(i,j)·
\frac{y(i)}{(R^n_i+S_i)}}}{\displaystyle\sum_{i\in Subset S}{A(i,j)·
\left(1+Penalty(j,n)\right)}}\right] \label{eq:eq3}
\end{equation}

Where the parameters and functions are defined as follows: 

\noindent
$x(j)$  - Activity of voxel $j$ ($j=1$, max. voxel number V)\\
$x^{n,s}(j)$ - Expected value of voxel $j$ at iteration $n$ and subiteration $s$\\
$A(i,j)$ - Probability that a photon emitted from voxel $j$ is detected at LOR $i$\\
$y(i)$ - Projection from the object measured at LOR $i$ (Experimental Data)\\
$S_i$ - Object Scatter and Random coincidences at LOR $i$ \\
$Penalty(j,n)$ - Penalty value at voxel $j$ and iteration $n$\\
$R_i^n$ - Projection estimated for the image reconstructed at iteration $n$:\\

\begin{equation}
R^n_i=\sum^{Max.\,\,\, voxel\,\,\, number}_{j=1}{A(i,j)x^n(j)} \label{eq:eq4}
\end{equation}

This MAP-OSEM algorithm can be considered as a generalization of the ML-EM. It incorporates a penalty 
MAP function which can be chosen in  different ways \cite{Stayman,Fessler,Nuyts}, and  
scatter and random counts estimates that may require additional modeling of these processes. 
OSEM reconstruction  without MAP regularization is obtained by setting the penalty function to zero.
We note, however, that in this work we are mostly interested in the way we compress the SRM and not in the effect of MAP on 
the image quality, and thus all the reconstructed images we present are obtained with zero penalty.

With regards to the number of iterations and subsets, 
we have found that reconstruction with 25+25+50 subsets exhibit a good compromise
between  resolution and reconstruction time. 
Thus, all the reconstructions presented in this work are obtained with three 
iterations of 25, 25 and 50 subsets (25+25+50) respectively.

The effect of scatter inside the object in 3D acquisitions of 
small animal has been studied recently
\cite{Yang2004,Yang2006}. It has been measured  a non negligible fraction of  recorded events 
coming from scatter. 
%
An accurate modeling of scatter at the object during reconstruction may improve image quality. 
In this work, scatter inside the object has been estimated assuming an isotropic and homogeneous 
scatter distribution.

With regards to attenuation, as it is a relatively minor effect for small animal 
PET \cite{Rutao2005} and our
main aim is to test the adequacy of our compressed SRM and not the importance of attenuation, 
we have not included it in the reconstructed images shown.

\section{SIMULATION RESULTS}

\subsection{Test set}
To test our methodology, we first reconstructed data from different simulated phantoms: 
uniform cylinders
and point sources in different axial and transaxial positions and simulated microresolution and Defrise 
phantoms. In order to study the linearity of the reconstruction method as well as the conservation 
of the number of counts and noise properties, a ``Spiral phantom'' was designed  (Fig. 2-3). It is comprised
of three cylinders (background) of 11.5 mm in diameter, each with two spirals inside: a hot one (activity 4 times greater 
than the background) and a cold one (activity 4 times lower than the background). The 
diameter of these
spiral shape cylinders are 1.4, 2.2 and 2.6 mm.
Events were generated from these test sets using our own MC method, taking into account positron range and 
non-colinearity effects.  Neither attenuation nor scatter within the object were included for these
simulations. The response 
of the detector was also realistically simulated considering the main physical effects contributing 
to the spread of the energy among crystals due to scatter in the scintillators. For each study, 
10 billion events were simulated and stored as projection data. We realized that the realistic model 
of detector response resulted in wider CHORs, which contained many more voxels than when more 
simplistic models of the system response are used. The images reconstructed from these simulations have a 
resolution of 175 $\times$ 175 $\times$ 62 voxels. The size of the phantoms and the images were chosen 
to be the same as the FOV of the eXplore Vista-DR (GE), namely 65 mm in diameter. Thus the voxel size 
is 0.38$\times$ 0.38 $\times $ 0.78 mm$^3$.

\subsection{Evaluation}
\begin{figure}
\begin{center}
\scalebox{0.8}{\includegraphics{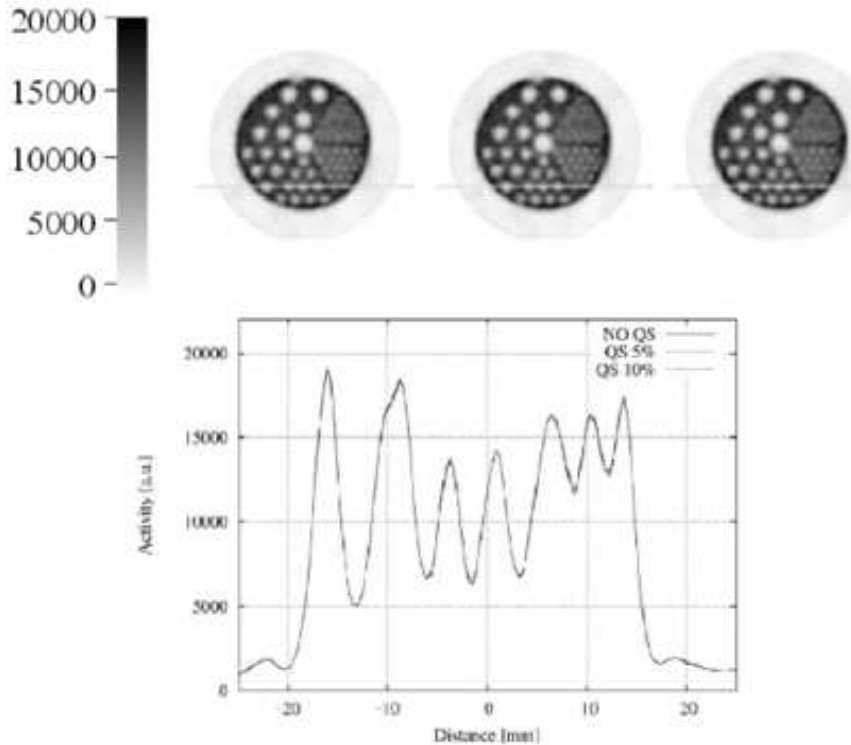}}
\caption{Reconstructions with different degree of quasi-symmetry assumptions.
Transaxial slices of a 25+25+50 3D-OSEM reconstruction of a cold Derenzo phantom (1 mCi of $^{68}Ge$, 90
minutes acquisition time) are shown at the top panel of the figure. Left slice is obtained without 
quasi-symmetries. The center one has been  reconstructed using the quasi-symmetries explained in 
Figures \ref{fig:chords3}, \ref{fig:chords_long} and
\ref{fig:chords_trans}. With this degree of  quasi-symmetry,
differences of less than 5\% inside superCHORs are found. 
The right slice has been obtained with a higher degree of quasi-symmetries (and compression), 
meaning up to 10\% of difference of the profiles inside a superCHOR. In the bottom panel of 
this figure, the activity profile against the distance (in mm) from the center 
of the line drawn in the slices is shown. Small differences in the
activity profile begin to be visible at the highest level of quasi-symmetry. Horizontal scale in mm.
Darker grey in the figure corresponds to larger activity values.
} \label{fig:derenzos}
\end{center}
\end{figure}

Initial tests were run to verify that the compressed SRM and the uncompressed SRM yielded images of
 similar quality and with no artifacts (see Figure \ref{fig:derenzos}). We will coment in more detail on  the effect of
compression in the SRM in the reconstructed images in next section.

 In a second step, an estimate of the Point Spread Function (PSF) 
was obtained by using a phantom consisting of an array of small sources located at different radial
 and axial positions and separated by 5 mm. Resolution obtained from reconstructions of simulated
 projections as well as from real phantoms are shown in Table \ref{tb:tb2}, revealing that submillimeter resolution
 can be obtained from real projections. As shown in the figures and summarized in Table \ref{tb:tb2}, very uniform 
values of resolution (as measured by FWHM) throughout the FOV of 0.7 mm (at center of the scanner) to 
0.9 mm (2.5 cm off axis) were obtained.  

\begin{table}
\caption{\label{tb:tb2}SPATIAL RESOLUTION}
\begin{indented}
\item[]\begin{tabular}{@{}*{7}{l}}

\br
Measured&R = 0 mm &R = 25 mm\\
Radial&0.7 mm&0.9 mm\\
Tangential&0.7 mm&0.9 mm\\
\br
Simulated (MC)&R = 0 mm&R = 25 mm\\
Radial&0.6 mm&0.8 mm\\
Tangential&0.6 mm&0.8 mm\\
\br

\end{tabular}
\end{indented}
\end{table}

\begin{figure}
\begin{center}
\scalebox{0.65}{\includegraphics{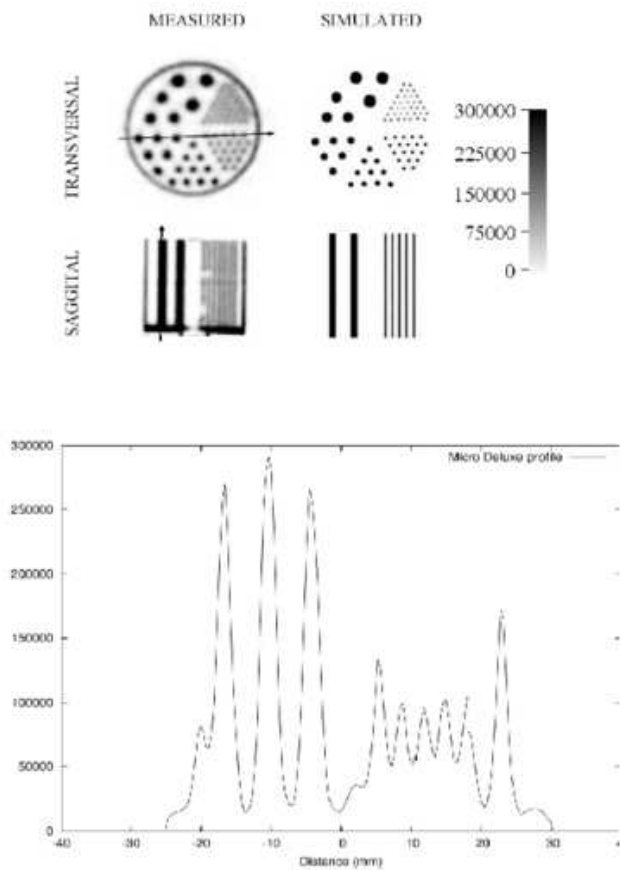}}
\caption{Micro resolution phantom, Data Spectrum Co., Hillsborough, NC.  (Top) Image reconstructed 
from real measured projections and from projections obtained after a simulation. Transversal
 and saggital views. Rod diameters 1.2, 1.6, 2.4, 3.2, 4.0 and 4.8 mm. Separation = twice diameter. 
(Bottom) Transverse line profile of the measured micro resolution phantom along the arrow 
indicated in the figure is shown.} \label{fig:diag4}
\end{center}
\end{figure}

These results of resolution can also be observed with the Micro resolution phantom reconstruction displayed
in Figure \ref{fig:diag4}, where the uniform resolution, almost constant along the radial direction, can be observed.

\begin{figure}
\begin{center}
\scalebox{0.5}{\includegraphics{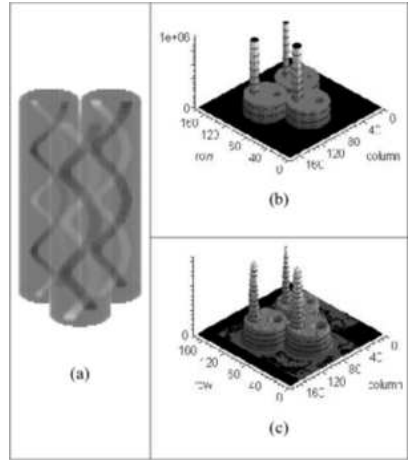}}
\caption{(a) Spiral-Phantom (b) 3D representation of a transverse section of the original 
Spiral-Phantom. 
Z-Axis represents counts. (c) 3D representation of the OSEM-3D reconstructed image after 3 iterations 
[25 + 25 + 50 subsets]. The three large cylinders are 11.5 mm in
size, and there are two small ones inside each large cylinder, a hot one,
with activity 4 times larger than the average on the large cylinder, and a cold one,
with activity 4 times smaller than the average one. Small cylinders are 1.4, 2.2 and 2.6 mm in diameter. } \label{fig:diag2}
\end{center}
\end{figure}

\begin{figure}
\begin{center}
\scalebox{0.6}{\includegraphics{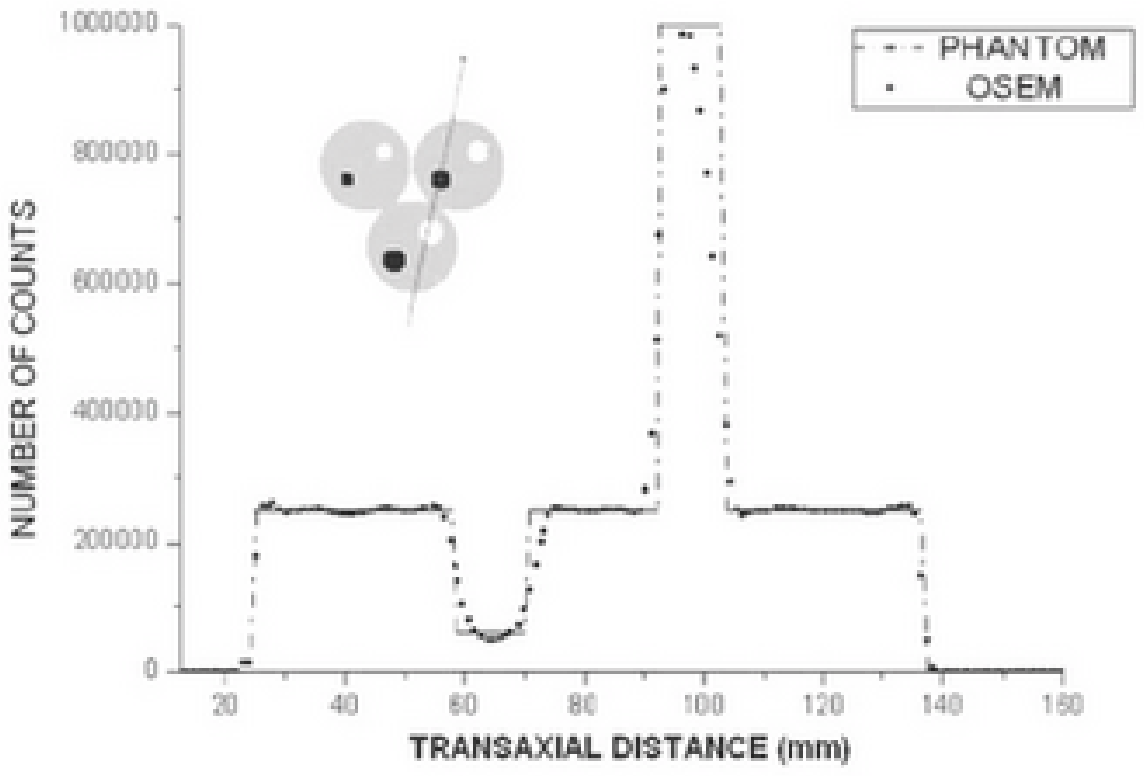}}
\caption{Profiles across the Spiral-phantom study showing the activity distribution:  
Phantom (solid line); OSEM reconstruction (dots) [25+25+50 subsets]. Voxel size is 
$0.38 (X) \times 0.38 (Y) \times 0.78 (Z)$ mm$^3$. } \label{fig:diag3}
\end{center}
\end{figure}

With regard to linearity, Figures \ref{fig:diag2} and 
\ref{fig:diag3} show an
 Spiral-Phantom and the reconstructed (OSEM) image after three iterations of
25+25+50 subsets. Note the very linear response exhibited by the reconstructed image: 
The hot spiral to background and 
background to cold spiral  activity ratios are preserved after reconstruction.

\section{EVALUATION OF THE EFFECT OF COMPRESSION}

To study the effect of the quasi-symmetries we have implemented, 
we have chosen a 90 minutes acquisition of a Cold Derenzo phantom of 1 mCi activity of $^{68}Ge$.
In Figure \ref{fig:derenzos} we show an slice of the phantom reconstructed after three 3D-OSEM iterations of
25+25+50 subsets were the SRM was dealt for in three ways:
a) Without making use of the quasi-symmetries. 
b) With the quasi-symmetries 
explained in previous sections, and  quasi-symmetry classes (superCHORs) built from  profiles that typically
differ by less than 5\%.
This allows for a reduction in a factor of approximately 9 in the
size of the SRM that needs to be stored. c) With a larger degree of compression, which allows for a
reduction factor in size of the SRM of approximately 25, with superCHORs that represent profiles that
differ approximately by less than 10\%.
In the bottom part of the figure we show the activity
profiles along the  lines indicated in each slice of the upper part of the figure. While the activity
profile of the reconstruction obtained without quasi-symmetry (solid line) and the one of the reconstruction
obtained with moderate compression (labeled QS 5\%, medium dashed line) are hardly distinguishable, the
reconstruction obtained with the most compressed SRM (labeled QS 10\%, short dashed line) begins to deviate
slightly from the uncompressed result.

Apart from this figure where we have studied the effect of quasi-symmetries, in the remaining of this work we
have employed the moderate amount of quasi-symmetries, which implies for the Vista DRT scanner an SRM size of
150 MB.

\section{RESULTS FROM SMALL ANIMAL DATA}

\begin{figure}
\begin{center}
\scalebox{0.6}{\includegraphics{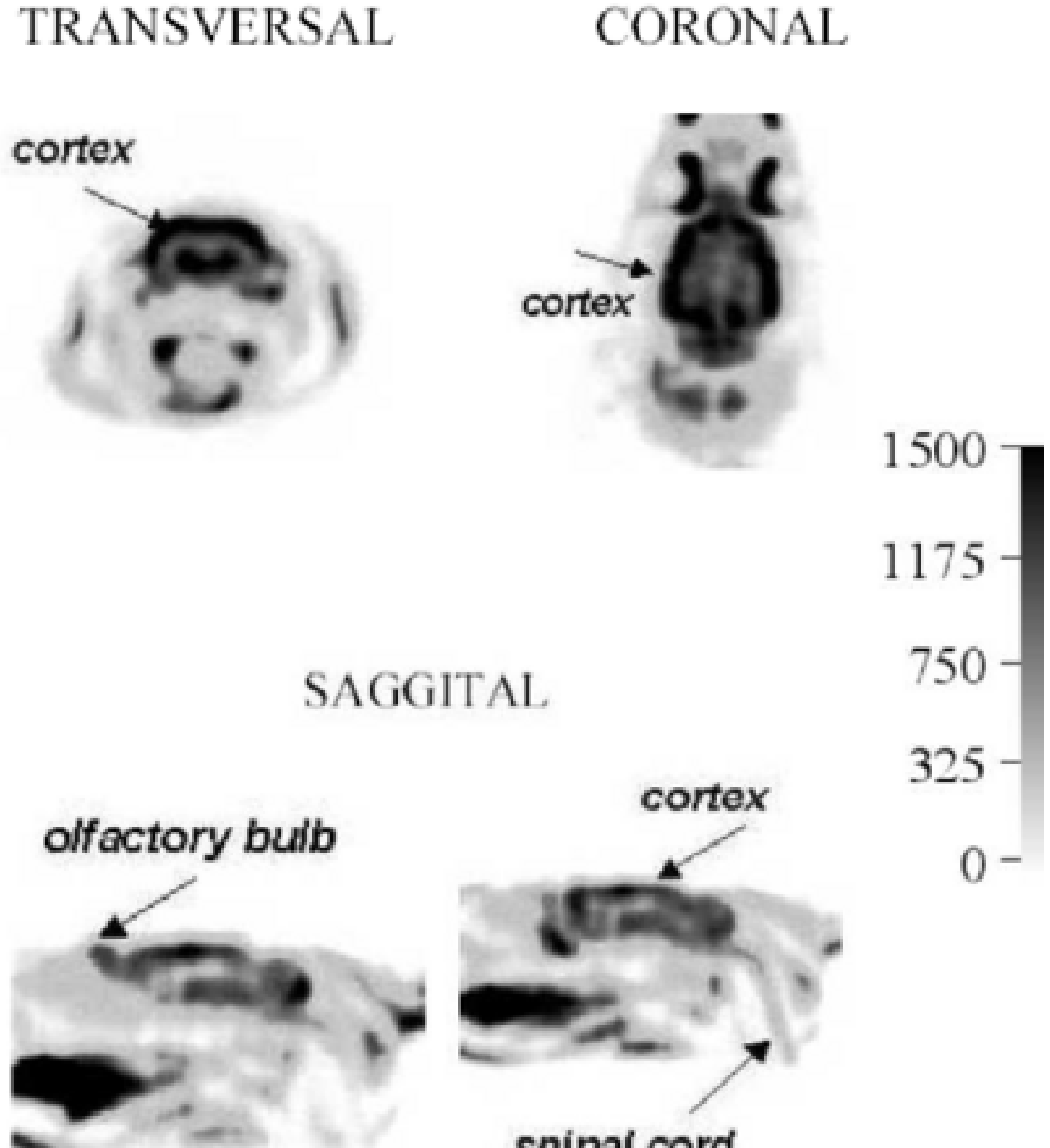}}
\caption{Single bed study of the head of a 185 g rat. 35 minutes intake of 1 mCi of FDG and 60 minutes
 scan acquisition in an eXplore Vista (GE) dRT PET scanner. 3D OSEM with 3 iterations 
of 25+25+50 subsets was employed.} \label{fig:diag5}
\end{center}
\end{figure}

\begin{figure}
\begin{center}
\scalebox{0.6}{\includegraphics{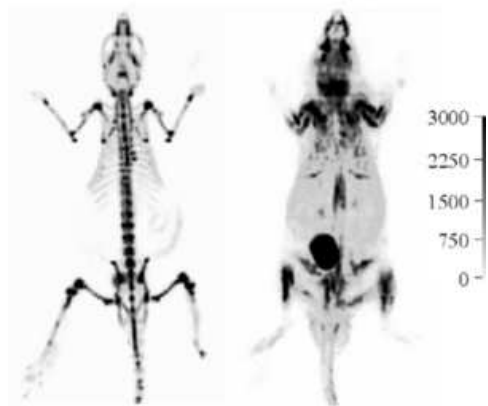}}
\caption{Reconstructed images of (a) $^{18}F$ and (b) FDG mouse study acquired with an eXplore 
Vista (GE) drT PET scanner. A three-bed scan of a 25 g mouse, with 5 minutes scan acquisition per bed, 
4 slices overlap between beds, 45  minutes uptake of 250 $\mu$Ci of FDG (right) or $^{18}F$ (left). 3D-OSEM 
with 3 iterations of 25+25+50 subsets were employed in both cases.} \label{fig:diag6}
\end{center}
\end{figure}

Our reconstruction software was also tested on real mice data. $^{18}F$ and FDG mice whole-body projections 
were acquired with a VISTA (GE) dRT PET scanner \cite{Vaquero1}. 
Figures \ref{fig:diag5} and \ref{fig:diag6}  show
 the reconstructed images obtained using the 3D-OSEM algorithm with 3 Iterations of 25+25+50 Subsets. 
The number of voxels is
 175$\times$175$\times$62 for the rat head depicted in
in Figure \ref{fig:diag5}, and 175$\times$175$\times$168 for the whole body
(three beds) mice of Figure \ref{fig:diag6}. In all cases, the voxel size is 0.38 $\times$ 0.38 $\times$ 0.78 mm$^3$. 
As indicated by the study from phantoms and simulated data, submillimetric
 details can be observed in the images of the mice and the rat head. 
In the rat head, cortex, spinal cord and olfactory bulb are easily identified. For the mice results,
the fluorhine image clearly shows small details such as ribs and spinal bones, and the small bones 
in the front
legs. The FDG image shows the usual accumulation of activity at the mouse urinary bladder, but no artifacts
are produced in its vicinity.

\section{PERFORMANCE ANALYSIS}

\subsection{Optimization techniques during 3D-OSEM reconstruction}

Considering all the strategies described previously, we implemented an accelerated 
version of OSEM that can optionally incorporate a penalty function in the reconstruction process 
(MAP-OSEM).  The number of subsets in each iteration can be chosen freely in between 1 and 100, 
not limited by system symmetries. Subsetting strategies require that all the subsets have
CHORs evenly distributed along the FOV. In order to achieve this, we pick superCHORs in  
random order and assign them consecutively to each subset. As all the CHORS belonging to the 
a superCHOR lie within the same subset, we can take advantage of symmetries and 
quasi-symmetries to speed up decompression of the SRM, because every CHOR needs to be decompressed once and
only once during  each iteration.
Subsets are thus chosen so that they include {\em all} members of a quasi-symmetry class.
 From 1 to 100 subsets, there are plenty of choices to build the subsets that fulfil this requisite 
of including all the members of  the classes comprised in a subset. We just chose symmetry classes
 in each subset at random.  Beyond 100 subsets, however, problems arise because every subset 
will then  include too few symmetry classes.

Always bearing in mind flexibility as a goal of design,  the number of subsets as well as the 
number and size of voxels or the size of the FOV employed can be changed at any time during 
reconstruction, even before full iterations are completed. 
In this way, we can try different sizes of  voxel, number of subsets and iterations and look for the 
best combination in terms of speed, quality of the reconstruction, or both.  For instance, 
during the first stages of reconstruction, when only the low frequencies (gross details) of the image 
are recovered, the use of a high number of voxels is a waste of computer power. The number
of voxels employed to represent the image may be increased as the high frequency components
 of the reconstructed image begin to appear. This strategy has been described in detail in \citeasnoun{Raheja}
 and it has been named {\em multiresolution}. We note that the images and execution times quoted
in present work have been obtained without resorting to this feature.

With regards to reconstruction times, a full iteration of an acquisition
covering the whole axial FOV (that is, an acquisition of ``one bed'' or 175$\times$175$\times$62 voxels) typically takes 30 minutes using 1 CPU 
(Opteron 244, 1800 GHz, 2GB RAM). Thus 90 minutes are needed for the single-bed, 
three-iteration reconstructions shown in Figure \ref{fig:diag5} of this work in a single CPU computer. For larger animal like rats that span a length larger than the
axial FOV of 4.8 cm, the bed that supports the animal is displaced during acquisition and thus 
several scans (bed positions) are acquired consecutively in order to cover the whole
body. More axial slices will be acquired and reconstruction time will 
be increased proportionally to the size of the axial FOV.

Reconstruction time scales approximately with the product of the number of LORs (2.9$\times10^7$) 
and the number of voxels in a LOR (on average 6000 for the standard resolution employed of 
175$\times$175$\times$62 voxels). 
Without compression, a similar reconstruction  needs to access above 3 GB worth of SRM elements
 from disk for every subset, which slows downs the reconstruction by a
 factor of 10 to 50, 
depending on disk speed and network activity.

The reduction of the storage needs for the SRM, allows to keep it in RAM. The code 
is implemented in a way that no disk I/O is needed in order to forward and backward project. 
The SRM is read at once at the beginning of the execution and the image is written to disk only 
after a full iteration is finished. Except for the short initial and final periods of intense 
disk I/O, the common Unix tools  measure a CPU use larger than 99\%, which indicates that the elapsed 
time during execution is mostly CPU bounded. Determining the performance of computer codes, however,
is a very subtle and non trivial task. A program can be 
 CPU bounded yet it may be wasting CPU cycles doing nothing useful.
 
CPU manufacturers often quote  peak performance of modern CPUs,  referring to ideal situations
where no cache misses occur,  burst mode acces to memory is possible, the CPU internal
pipelines are fully used, etc. For instance, a peak performance of 2 flops per cpu cycle is
quoted for AMD Opteron CPUs, what refers to a single multiply and add instruction performed
in a clock cycle.
 Real life applications depart from
the ideal conditions and thus peak CPU performance is hardly achieved during sustained execution 
of complex codes.  
In order to assess the perfomance of the code, 
FIRST was compiled with the 8.1 version of the Intel fortran compiler. 
The Intel vtune performance analysis and profiling 
tool was employed to determine performance and number of flops required by each routine.
We conclude that during reconstruction, 50\% of peak performance
was obtained in  sustained fashion. We also determined that 
when our compressed SRM 
 that fits in RAM memory is employed, the decompression time measured was in the range of 10\% to 30\%
of total reconstruction time, depending on the number of voxels chosen. 

\subsection{Parallel implementation}

Parallel computing on multiple processors is an attractive option in order to reduce 
computational time. The use of protocols like the Message Passing Interface (MPI) enables 
clusters of networked industry-standard PC's (Beowulf clusters)  
to be relatively easy configured as a multiprocessor unit. 
Several parallel implementations of the OSEM algorithm have been presented in other
works \cite{Calvin,Calvin2,Chen98}. We have decided to implement 
a parallel version of our Fast Iterative Reconstruction 
Software (FIRST) to run in  Beowulf clusters of several CPUs in a master/slave 
configuration, characterized by the use of a master-process and several (usually as many 
as the number of available CPUs) slave-processes. The master distributes the job among the 
slaves and continuously balances the workload, to achieve the best performance taking into 
account differences in individual speed or workload on each CPU. In spite of its name, the
master process does not perform any intensive calculations, though. On startup, the master 
process reads from disk (once and for all) the SRM elements and sends them to the slave processes.
Enough RAM memory to store the full compressed SRM must be available for each slave process. After 
startup, the master process decides which part of each subset (i.e., which actual superCHORs of such
subset) is forward and backward projected by each slave process. Once all the slave process
have finished with their share of each subset, the master process updates the image, that is also
stored in memory, and broadcasts the new image to all the slave processes. Upon completion of the 
reconstruction, the master process updates the image on disk. The slave processes
are highly CPU-intensive, as they are continuoulsy performing the forward and backward
projection operations. 

The master process only  takes part in the reconstruction whenever
one of the slave processes finishes its share of the reconstruction task and claims for more, or
when all the CHORs of a subset have been visited by all the slave processes and then the image must be 
updated and broadcasted. Multitask capabilities of modern computers and operating systems makes
it possible to have as many slave processes as available CPUs, yet having an additional master
process that will occupy a few cycles of one of the CPUs that is running one of the slave
process. Balance of the workload among different CPUs is easily achieved as the slave processes that run faster
(because they are executed by a less busy or faster CPU) will claim for their share of subsets
more often than the ones that run slower. The only caution that must be taken is that the
initial workload sent to each slave process is similar but not identical for all of them. In this way,
we  minimize the possibility that more than one slave process claims the attention of
the master process at the same time. In practice, the tasks assigned to the slave process take 
from a few seconds to near a few tenths of seconds to be completed before requesting the action
of the master process. The master process, on the other hand, can comply with the task required by
the slave process in just a fraction of a second. 

\begin{table}
\caption[caption help]{\label{tb:tb3}FIRST RECONSTRUCTION ELAPSED TIME\\
Elapsed times for reconstruction of 50 subsets for 1 bed reconstructed with $ 175 \times 175 \times 62 $
voxels are displayed on dedicated machines. The same code, compiled with
intel fortran compiler 8.1 and 32 bits libraries is employed in both systems. Fedora
Core 3.0 x86-i386 operating system. At least 1 GB per CPU is available in all cases. The 
elapsed time in minutes is shown. In both cases, the clusters were configured with dual nodes 
connected via  GB 
ethernet cards
and switches.}
\begin{tabular}{@{}*{7}{clll}}
\br
 CPUs & Version & CPU class & elapsed time (min.) \\
\hline
 1             & Non parallel & AMD Opteron 244 1.8 GHz &  32.3   \\
 1             & 1 Master + 1 Slave & {\em id.} & 35.2   \\
 2             & 1 Master + 2 Slaves & {\em id.}  &  18.2   \\
 4             & 1 Master + 4 Slaves & {\em id.} & 9.4    \\
 8             & 1 Master + 8 Slaves & {\em id.}  & 5.0    \\
\hline
 1             & Non parallel & Intel Xeon EMT64 2.8 GHz & 43.6    \\
 1             & 1 Master + 1 Slave & {\em id. }& 46.1    \\
 2             & 1 Master + 2 Slaves & {\em id.}  &  24.3   \\
 4             & 1 Master + 4 Slaves & {\em id.}  & 12.2    \\
\hline
\br
\end{tabular}
\end{table}

Overall, the implementation is simple and
efficient. In Table \ref{tb:tb3} we quote the elapsed time in minutes taken by the
reconstruction of one iteration of 50 subsets at nominal number of voxels
of $175 \times 175 \times 62 $. We made  tests in an one-CPU system comparing the 
parallel to the nonparallel versions. A  master plus a slave process parallel
reconstruction  in a single CPU takes less than $10$\% longer than the nonparallel 
(only one process in one CPU)
version of the code working in the same single-CPU system. The additional time is due to 
the overhead of sending the SRM from the master to the slave, 
as well as the elements of the image
 after updates, using the MPI interface.  On the other hand, the parallel version of FIRST 
working over several CPUs reduces the elapsed reconstruction time by a factor nearly equal to the number 
of CPUs available. For relatively small clusters with up to 8 CPUs, the results shown in the
table \ref{tb:tb3} indicate that the implementation  
with only one master process is rather efficient. When two or more CPUs are available, the master
process uses less than 2\% of the total computing time required for the reconstruction, 
according to the CPU use stated by the Unix common tools function  {\em ps}.

If a number of CPUs larger than 8 is to be used, benefits will be found by using more than one
master process. In Table \ref{tb:tb3} a summary of elapsed time for the same reconstruction 
over two different platforms using one master process is shown.

\section{DISCUSSION AND CONCLUSIONS}

We implemented FIRST, a fully 3D-OSEM or 3D-MAP-OSEM non-sinogram-based reconstruction algorithm, 
using a compressed SRM 
that contains the resolution recovery properties of EM. The full SRM can be stored in less than 150 MB
 of storage. Reconstructed images are indistinguishable from the ones obtained without compression. 
The use of the compressed SRM allowed for a reconstruction with  a more realistic response of
 the system. In this work, we used our own MC model of the scanner which incorporates physical effects 
such as positron range, non-colinearity and scatter in the scintillator material.  Although it took
 several weeks, the SRM was computed only once.  It was stored in compressed form so that the 
reconstruction program could keep it in dynamic memory. 

Thanks to this, near peak performance of the 
algorithm is achieved, with just a slight overhead (10-30\%) due to the decompression procedure. 
This resulted in short reconstruction times, even if the realistic SRM implies wider CHORS and thus
 more voxels are involved in every projection and back projection operation than when 
simplified SRM are used. The algorithm has been validated against simulations as well as 
real data. Acquisitions of phantoms and mice from a commercially available high resolution 
PET scanner were reconstructed. A realistic SRM from our own MC model of the scanner with optimal 
resolution recovery was used. This fact, together with the intrinsic high resolution 
(small crystal pitch of 1.55 mm) of the scanner, resulted in very high quality images with 
submillimeter resolution, as shown in Figures \ref{fig:diag4}, \ref{fig:diag5} and \ref{fig:diag6}.
 The reconstruction time needed by the algorithm enables real time operation in a small 
cluster (less than 10 minutes per bed and iteration in a 4 CPUs cluster) of industry-standard PC's. 
The results from real acquisitions in terms of  resolution and linearity agree with what is 
expected from the simulated projections that use the  same SRM as the reconstructions. 
This indicates that the SRM derived from our Monte Carlo simulations accurately reflects 
the response of the real scanner. Very uniform resolution and linearity is exhibited by the 
reconstructed images. 

The flexibility, reduced reconstruction time, accuracy and resolution of the resulting images prove 
that the methodologies used to implement the FIRST reconstruction can be applied to real studies of
 high resolution small animal PET scanners. The use of quasi-symmetries to reduce (compress) the 
size of the SRM seem to be an adequate way of dealing with the problem of storing the huge SRM
 resulting from modern high resolution PET scanners.

\ack

The authors thank Dr. Martin G. Pomper, James Fox and J\"urgen Seidel from the Johns Hopkins University 
School of Medicine for providing access to the eXplore VISTA datasets. Part of this work is funded by
 the IM3 network (Ministerio de Sanidad), with grants from the Ministerio de Educaci\'on y Ciencia,
 projects BFM2003-04147-C02-01 and TEC2004-07052-C02-01, Fondo de Investigaciones Sanitarias
del Instituto de Salud Carlos III project PI052583 and Ministerio de Industria, 
 Turismo y Comercio projects FIT-330101-2004-3 and  CDTEAM (CENIT).

\References
\bibliography{first}
\endrefs

\end{document}